\documentstyle[twoside,11pt,draft]{article}
\newcommand{\be}{\begin{equation}}
\newcommand{\ee}{\end{equation}}
\newcommand{\bra}[1]{\mbox{$\langle\, #1 \mid$}}

\newcommand{\ket}[1]{\mbox{$\mid #1\,\rangle$}}

\renewcommand{\a}{\hat a}
\newcommand{\ac}{\hat a^{i \,\dagger}}
\renewcommand{\b}{\hat b}
\newcommand{\bc}{\hat b^{i \,\dagger}}
\newcommand{\fip}{\phi_{\bf k}^i}
\newcommand{\pip}{\pi_{\bf k}^i}
\newcommand{\g}{\Gamma (\nu)}
\newcommand{\argom}{( \Theta_k - \delta)}
\newcommand{\kap}{\left( \frac{k}{2H} \right)^{-\nu}}
\newcommand{\quadcap}{\left( \frac{k}{2H} \right)^{-2\nu}}
\newcommand{\kbig}{\vec{k}}
\newcommand{\xbig}{\vec{x}}
\title{Adiabatic Invariants and Scalar Fields in  a de Sitter Space-Time}
\author{C. Bertoni\thanks{e--mail: bertoni@astbo1.bo.cnr.it} \\
{\small\it Istituto di Radioastronomia, via Gobetti 101, 40129 Bologna Italy}
\and F. Finelli\thanks{e--mail: finellif@bo.infn.it}, G. Venturi 
\thanks{e--mail: armitage@bo.infn.it}\\
{\small\it Dipartimento di Fisica, Universit\`a degli Studi di Bologna
and I.N.F.N.} \\
{\small\it via Irnerio, 46 -- 40126 Bologna -- Italy}}

\begin{document}

\baselineskip 4.0ex
\begin{titlepage}
\pagestyle{empty}
\maketitle
\begin{abstract}
The method of adiabatic invariants for time dependent Hamiltonians is applied 
to a massive scalar field in a de Sitter space-time. The scalar field ground 
state, its Fock space and coherent states are constructed  and related 
to the particle states. Diverse quantities of physical interest 
are illustrated, such as particle creation and the way a classical 
probability distribution emerges for the system at late times.   
\end{abstract}

\end{titlepage}
\pagestyle{plain}
\raggedbottom
\setcounter{page}{1}

In a previous paper \cite{bertoni} we illustrated the Born-Oppenheimer (BO) 
approach 
to the matter-gravity system in a simple minisuperspace model and in such 
a context the simplifications associated with 
the use of adiabatic invariants \cite{lewis} were pointed out.
Indeed through the use of such invariants one can improve on the adiabatic 
approximation and obtain better results for the evaluation of 
fluctuations which, in a quantum gravitational context, are associated with 
the creation of matter. 

The purpose of this note is to illustrate, within the context of field theory 
in a curved space-time (de Sitter with flat spatial section 
in our simplified 
model), the use of adiabatic invariants with a particular emphasis on the 
description of the vacuum and the space of physical states.
The usefulness of the method of invariants for the calculation of the 
geometrical (adiabatic) and dynamical phases has been previously noticed 
\cite{gao} and the method of invariants itself applied to quantum cosmology 
\cite{abe}, although not in a BO context. Here we shall 
apply the method to the equation for matter obtained \cite{bertoni} in 
the BO approach, on neglecting fluctuations and in the semiclassical limit 
for gravity. This novel application allows for the time
variation of the metric in a second quantized (Schr\"odinger functional
approach \cite{guven} \cite{cinesi}) scalar field theory thus improving 
on the adiabatic approximation
(static metric) and including some matter creation through the use
of the vacuum and the Fock space associated with quantum adiabatic
invariants.

Let us consider a Friedmann-Robertson-Walker line element:

\be
ds^2 = - d\tau^2 + a^2(\tau) g_{ij} dx^i dx^j 
\ee

\noindent
where $ g_{ij}$ is the three metric for a flat three-space. 
We further introduce a real massive scalar field $\Phi(\xbig,\tau)$ 
whose lagrangian density is given by:

\be
{\cal L} = -\frac{1}{2} g^{\mu \nu} \partial_{\mu} \Phi \partial_{\nu} \Phi 
- \frac{1}{2} \xi R \Phi^2 - \frac{1}{2} \mu^2 \Phi ^2
\ee

\noindent
with $R$ the Ricci scalar 
and decompose the scalar field on a complete basis $u_{\bf k}$ :

\be
\Phi\, (\xbig,\tau) = \sum_{\kbig} 
\left( a_{\bf k} u_{\bf k} + a^*_k 
u^*_{\bf k} \right) \,,
\ee 
with
\be
u_{k}(\xbig,\tau)=\frac{1}{\sqrt V}
\, e^{i \,\kbig \cdot \xbig} \phi_{k} (\tau) 
\equiv \frac{1}{\sqrt V}
\, e^{i \,\kbig \cdot \xbig} \frac{\phi_{k}^1 
+ i \phi_{k}^2}{\sqrt 2}
\,,
\label{fourier}
\ee
where we have separated $\phi_k$ into its real and imaginary parts, and 
we consider normalization to a finite volume V. 

From the above one obtains an action:

\begin{eqnarray}
S = \sum_{{\kbig},i} S_k^i = \frac{1}{2} \sum_{\kbig} \sum_{i=1,2} 
\int a^3\,d\tau \left( \dot{\phi}_k^{i\,2} - \omega_k^2 \phi_k^{i\,2} \right)
\,,
\label{action}
\end{eqnarray}
where we have introduced a frequency $\omega_k^2 = \frac{k^2}{a^2} + \mu^2 + 
\xi\,R$ and by the dot we denote a derivative with respect to 
the time $\tau$. From Eq.(\ref{action}) one sees that the modes 
$\kbig$, i are independent and may therefore be considered separately, 
hence we need only consider a given mode. 
One obtains, for a mode $\kbig$, i, a Hamiltonian:

\be
H_{k}^i = \frac{1}{2a^3} 
\left( \pi_k^{i\,2} + a^6 \omega_k^2 \phi_k^{i\,2} \right) \,,
\ee

\noindent
where $\pip=a^3 \dot{\phi}_k^i$ and a classical equation of motion for 
$\fip$: 

\be
\ddot{\phi}_k^i + 3\frac{\dot a}{a} \dot{\phi}_k^i + \omega_k^2 \fip = 0 \,.
\label{class}
\ee
In de Sitter space ($a=e^{H\tau}$ and $R=12 H^2$, where H is the time 
independent Hubble costant)
this equation can be rewritten in terms of the conformal time 
$\eta=-\frac{1}{H}e^{-H\tau}$ ($a=-\frac{1}{H\eta} \,, -\infty < \eta <0$) as:

\be
\left[ \frac{d^2}{d(k\eta)^2} + \frac{1}{k\eta} \frac{d}{d(k\eta)} 
+ 1 - \frac{\nu^2}{(k\eta)^2} \right]a^{\frac{3}{2}} \fip=0
\ee
which has as solutions the Bessel functions $J_{\nu}(k|\eta|)$ and 
$N_{\nu}(k|\eta|)$ (or 
equivalently the Hankel functions $H_{\nu}^{(1)}$, $H_{\nu}^{(2)}$ for 
$u_k$, $u_k^*$ respectively) with: 

\be
\nu^2 = \frac{9}{4} - \frac{\mu^2 + 12 \xi H^2}{H^2} \,.
\ee 

Further, canonical quantization ($\pip \rightarrow -i \hbar \frac{\partial}
{\partial \fip}$) leads to: 
\be
\hat{H}_k^i = \frac{1}{2a^3} \left( -\hbar^2 \frac{\partial^2}
{\partial \phi_k^{i\,2}} + a^6 \omega_k^2 \phi_k^{i\,2} \right)
\label{hk}
\ee
which of course depends explicitly on time and may be rewritten as: 

\be
\hat{H}_k^i = \hbar \omega_k \left(\, \a_k^{i\,\dagger} \, \hat a_k^i 
+ \frac{1}{2}\,\right)
\ee
with: 
\be
\begin{array}{c}
\hat{a}_k^i = \left( \frac{a^3 \omega_k}{2 \hbar} \right)^{\frac{1}{2}}
\left( \hat\fip + i\,{\hat\pip \over a^3\omega_k} \right) \,,
\\
\\
\ac_k = \left( \frac{a^3 \omega_k}{2 \hbar} \right)^{\frac{1}{2}}
\left( \hat\fip - i\,{\hat\pip \over a^3\omega_k} \right) 
\ .
\end{array}
\ee 
and $[\hat a_k^i, \ac_{k'}]=\delta_{k k'}$.

A suitable method for the study of time dependent quantum 
systems is that 
of adiabatic invariants \cite{lewis}. In particular a hermitian operator
$\hat I_{k}^i(\tau)$ is called an adiabatic invariant if it satisfies 
\cite{lewis}:

\be
{\partial \hat I_{k}^i(\tau)\over \partial\,\tau}
- \frac{i}{\hbar} [\hat I_{k}^i (\tau),\hat H_k^i(\tau)]=0 \,.
\ee
The adiabatic invariant $\hat I_k^i$ has real, time independent, eigenvalues 
and 
in our case, can be decomposed in terms of basic linear invariants \cite{gao}:

\be
\begin{array}{c}
\hat I_{b,k}^i(\tau) \equiv 
e^{i\Theta_k(\tau)} \b_k^i(\tau) \equiv { e^{i\Theta_k} \over \sqrt{2\,\hbar}}\,
\left[{\hat \fip \over \rho_k}+
i\,\left(\rho_k\,\hat \pip - a^3 \dot{\rho}_k \hat \fip \right)\right] \,,
\\
\\
\hat I_{b,k}^{i\, \dagger} (\tau) \equiv
e^{-i\Theta_k (\tau)} \bc_k(\tau) \equiv { e^{-i\Theta_k} \over 
\sqrt{2\,\hbar}}\,\left[{\hat \phi_k^i \over \rho_k}-
i\,\left(\rho_k\,\hat \pi_k - a^3 \dot{\rho}_k \hat \phi_k^i \right)\right]
\,,
\end{array}
\ee
where $\rho_k(\tau)$ is real (see however 
\cite{wollenberg}) and satisfies \cite{lewis}:

\be
\ddot{\rho}_k + 3 \frac{\dot a}{a} \dot{\rho}_k
+ \omega^2_k \rho_k = \frac{1}{a^6 \rho_k^3}
\label{rho}
\ee
with:
\be
\Theta_k(\tau)=\int_{-\infty}^\tau
{d\tau'\over a^3(\tau')\,\rho_k^2(\tau')} \,.
\ee
The quadratic, hermitian, adiabatic invariant originally 
introduced in \cite{lewis} is 
therefore:  

\be
\hat I_{k}^i (\tau) = 
\hbar \left(\bc_k \,\b_k^i + \frac{1}{2} \right) = \frac{1}{2}
\left[ \frac{\hat \phi_k^{i\,2}}{\rho_k^2} + (\rho_k \hat \pip - a^3 
\dot{\rho}_k \hat \fip)^2 \right]
\ ,
\ee
with $[\b_k^i,\bc_{k'}]=\delta_{k k'}$.
The existence of the destruction operator $\hat b_k$ allows us to introduce 
a vacuum state $\chi_0(\fip)=<\phi|\kbig,0,\tau>_i$ defined by:

\be
\b_k^i \chi_0(\fip) = \left[ \frac{\hat \fip}{\rho_k} +
i\,\left( \rho_k\,\hat \pip - a^3 \dot{\rho}_k \hat \fip \right)\right] 
\chi_0(\fip) = 0
\label{vuoto}
\ee
leading to a normalized wave function: 
\be
\chi_0(\fip) = \left( \frac{1}{\rho_k^2 
\hbar \pi} \right)^{\frac{1}{4}} \exp \left[-\left( 
\frac{1}{\rho_k^2} - 
i\, a^3\, \frac{\dot{\rho}_k}{\rho_k} \right) 
\frac{\phi_k^{i\,2}}{2 \hbar} \right] 
\label{chi1}
\ee
since $\fip$, $\rho_k$ are real. We further note that the general 
solution to (\ref{rho}) may be written in terms of two solution to (\ref
{class}) (say $a^{-\frac{3}{2}} J_\nu$, $a^{-\frac{3}{2}} N_\nu$ ) as
\cite{cole}:
\be
\rho_k = a^{-\frac{3}{2}}\left[ A J_\nu^2 + B N_\nu^2 + 2(AB - \frac{\pi^2}
{4 H^2})^{\frac{1}{2}} J_\nu N_\nu \right]^{\frac{1}{2}} \,,
\label{eqrho}
\ee
where A,B are real constants. It is clear that the vacuum state is then 
defined up to two real parameters (or one complex parameter) \cite{
bunch} \cite{guth}. Let us note that the above vacuum state is quite 
distinct from the harmonic oscillator one, indeed one has that:
\begin{eqnarray}
\hat a_k^i &=& \frac{1}{2} \b_k^i \left[ \rho_k (a^3\omega_k)^{\frac{1}{2}} + 
\frac{1}{ \rho_k (a^3\omega_k)^{\frac{1}{2}}} + i \dot{\rho}_k \left(\frac{a^3}
{\omega_k}\right)
^{\frac{1}{2}} \right] + \nonumber \\
&+& \frac{1}{2} \bc_k 
\left[ \rho_k (a^3\omega_k)^{\frac{1}{2}} - \frac{1}{\rho_k (a^3\omega_k)^
{\frac{1}{2}}}
+ i \dot{\rho}_k \left(\frac{a^3}{\omega_k}\right)
^{\frac{1}{2}} \right] \equiv \nonumber \\
&\equiv& \mu (\tau)\, \b_k^i + \nu (\tau)\,\bc_k \,, 
\label{bogo}
\end{eqnarray} 
and similarly for $\ac_k$ ,
that is the $\a_k^i$ and $\b_k^i$ (${\hat I}_{b,k}^i$) 
operators are related through a 
Bogolubov transformation \cite{gao}, and in particular one may relate 
them through a unitary transformation ({\sl squeezing} plus {\sl rotation}) 
\cite{squeeze}:
\be
\hat a_k^i = S(r,\varphi) \, R(\theta) \,\b_k^i\, R^\dagger (\theta) 
S^\dagger (r,\varphi) \,,
\ee
where the rotation operator R is: 
\be
R(\theta) = \exp (-i\theta \bc_k \b_k^i) \,,
\ee
and the squeezing operator S is: 
\be
S(r,\varphi) = \exp\left[ \frac{r}{2} \left( e^{-2i\varphi}\b_k^{i\,2} 
- e^{2i\varphi}\b_k^{i\,\dagger\,2} \right)
\right] \,,
\ee
with: 
\be
\left\{\begin{array}{l}
\tanh \, r = \left| \frac{\nu}{\mu} \right|
\\
\tan \, \theta = \frac{Im \mu}{ Re \mu}
\\
\tan (2\varphi - \theta) = \frac{Im \nu}{ Re \nu} \,.
\end{array}\right.
\ee
Let us remember that such a Bogolubov transformation is not unitarily 
implementable if one consider all modes with an infinite volume V 
\cite{umezawa}. Further we note that the above transformation is 
between two different Fock space basis at equal times and not between 
the same basis at different times.

The two different vacua will coincide in the adiabatic limit (costant 
frequency \cite{lewis}) for which 
the derivatives of $\rho_k \, a^{\frac{3}{2}} $ are small. 
One then has from (\ref{rho}):
\be
\rho^4_k \simeq \frac{1}{a^6 \omega_k^2} = 
\frac{1}{a^4} \left[ k^2 + a^2 \left( \mu^2 +
\xi R \right) \right] ^{-1}
\ee
in which case $\a_k$ and $\b_k$ coincide and: 
\be
\omega_k \hat I_k^i (\tau) \simeq \hat H_k^i (\tau) \,.
\label{limad}
\ee
This will occur for very early times ($\tau \rightarrow - \infty$) or 
for wavelengths $\frac{2 \pi a}{k}$ which are very small compared to 
the de Sitter horizon $H^{-1}$. In such a limit ($-k\eta = \frac{k}{H}
e^{-H\tau} \rightarrow \infty$) Eq.~(\ref{eqrho}) leads to (with 
$\Lambda = \frac{k}{H} e^{-H\tau}
-\frac{\pi}{2}\nu -\frac{\pi}{4}$):

\be
\frac{1}{k} = \frac{2H}{\pi k} \left( A \cos^2 \, \Lambda 
+ B \sin^2 \, \Lambda + 2 \left( AB - \frac{\pi^2}{4 H^2}
\right)^{\frac{1}{2}} \sin \, \Lambda \cos \, \Lambda \right)\,,
\ee
which has as solution: 
\be
\begin{array}{c}
A=B=\frac{\pi}{2 H}
\\
\\
\rho_{ok} = a^{-\frac{3}{2}} \left( \frac{\pi}{2H} \right)^{\frac{1}{2}}
(J_\nu^2 + N_\nu^2)^{\frac{1}{2}}
\label{Bunch}
\end{array}
\ee 
whereupon eq. (\ref{chi1}) becomes: 
\be
\chi_0(\fip) = \left( \frac{1}{\rho_{0k}^2 \hbar \pi} 
\right)^{\frac{1}{4}} \exp \left[
-\left( \frac{1}{\rho_{0k}^2} -
i\, a^3\, \frac{\dot{\rho}_{0k}}{\rho_{0k}} \right) \frac{\phi_k^{i\,2}}
{2 \hbar}\right]\,.
\label{bunch}
\ee

The expressions we have obtained for the vacuum in Eqs.(\ref{chi1}),
(\ref{bunch}), although 
apparently different, are actually the same as those previously obtained \cite
{guth}. Indeed one may write:
\be
\frac{1}{\rho_k^2} - i a^3 \frac{\dot{\rho_k}}{\rho_k} 
= -i a^3 \frac{\dot{\psi}_k}{\psi_k} 
+ \frac{3}{2} ia^2 \dot a
\ee
where:
\be
\frac{\psi_k'}{\psi_k} = \frac{ \frac{1}{2} \frac{d}{d\tau}  \left\{
\left[ J_\nu + \frac{1}{A}(AB - \frac{\pi^2}{4 H^2})^{\frac{1}{2}} N_\nu
\right]^2 + \frac{N_\nu^2 \pi^2}{4 A^2 H^2} \right\} + \frac{i}{A}}
{\left[ J_\nu + \frac{1}{A}(AB - \frac{\pi^2}{4 H^2})^{\frac{1}{2}} N_\nu
\right]^2 + \frac{N_\nu^2 \pi^2}{4 A^2 H^2} }
\ee
which for the solution Eq.~(\ref{Bunch}) becomes: 
\be
\frac{\dot{\psi}_k}{\psi_k} = \frac{ \dot J_\nu J_\nu + \dot N_\nu N_\nu 
+ \frac{2i H}{\pi}}
{J_\nu^2 + N_\nu^2} = \frac{\dot J_\nu - i \dot N_\nu}{J_\nu - iN_\nu} = 
\frac{\dot H_\nu^{(2)}}{H_\nu^{(2)}} \,.
\ee
The above result actually corresponds to the usual Bunch-Davies \cite{bunch}
vacuum in the 
Schr\"odinger picture \cite{guth} and we have chosen it among the diverse de 
Sitter invariant vacua by implementing adiabaticity for very early times.

The general vacuum state (\ref{vuoto}) 
which satisfies the Schr\"odinger equation associated with 
$H_k^i$ may be written as \cite{lewis} \cite{gao}:
\be
|\kbig,0,\tau>_{i,s} = e^{- i \frac{\Theta_k}{2}} 
|\kbig,0,\tau>_i
\ee
and we may construct arbitrary eigenstates $|\kbig,n,\tau>_{i,s}$: 
\be
\ket{\kbig, n, \tau}_{i,s} = e^{-i\,n\,\Theta_k(\tau)}\,
{(\bc)^n\over\sqrt{n!}}\,\ket{\kbig,0,\tau}_{i,s}
\ee
and a coherent state \cite{hartley} \cite{pedrosa}:
\be
\ket{\kbig, \alpha, \tau}_{i,s} = e^{-|\alpha|^2/2}\,
\sum\limits_{n=0}^\infty\,{\alpha^n\over\sqrt{n!}}\,
\ket{\kbig, n,\tau}_{i,s}
\ee
where $\alpha=u+i\,v$ is an arbitrary constant, both of which again satisfy 
the Schr\"odinger equation associated with $\hat H_k^i$ (\ref{hk}). Let us 
emphasize that our construction of the above Fock space and the related 
coherent states through the use of adiabatic invariants generalises 
previous work in which only the vacuum state was analyzed \cite{guven} 
\cite{guth}.  
 
Given the coherent state one may evaluate 
(with $<O>_s \equiv  \,_{i,s}\bra{\kbig, 
\alpha,\tau}\,\hat O \, \ket{\kbig, \alpha,\tau}_{i,s}$) :
\be
<\fip>_s 
=\sqrt{2\,\hbar}\,|\alpha|\, \rho_k \,\cos(\Theta_k-\delta) \,
\label{ficlass}
\ee
with $\tan\,\delta = \frac{v}{u}$,
\be
<\pip>_s= \sqrt{2\,\hbar} \, |\alpha|\,\left[
a^3 \,\dot{\rho}_k \,\cos(\Theta_k -\delta)
-{1\over \rho_k}\,\sin(\Theta_k -\delta)\right]
\label{piclass}
\ee
\be
<(\Delta\fip)^2 >_s \equiv
< \,(\hat \fip - <\fip>_s)^2\, >_s
={\hbar\over 2}\,\rho_k^2  
\ee
\be
<(\Delta\pip)^2>_s \, \equiv\ \,
<(\hat \pip - <\pip>_s)^2\,>_{s}
={\hbar\over 2}\,\left({1\over \rho_k^2} + a^6\,\dot{\rho}_k^2 \right)
\ ,
\ee
where $<\fip>_s$ can be verified to be a solution of eq.(\ref{class}). 
Thus through the use of adiabatic invariants we obtain solutions 
to the classical equations of motion and therefore a correct definitions 
of physical states \cite{fulling}, which are related to the particle states 
(associated with $\hat a,\, \hat a^\dagger$) through a Bogolubov 
transformation.

Further from the above one obtains:

\be
< \hat I_k^i >_s = \hbar \left( |\alpha |^2 + \frac{1}{2} \right)
\label{inv}
\ee

\be
<(\Delta\fip)^2>_s^{\frac{1}{2}} \,<(\Delta\pip)^2>_s^{\frac{1}{2}}=
{\hbar\over 2}\,\sqrt{1+ a^6 \,\rho_k^2\,\dot{\rho}_k^2}
\ ,
\label{indeter}
\ee

\begin{eqnarray}
<H_k^i>_s &=& \frac{1}{2a^3} \left[ <\hat \pi_k^{i\,2}>_s + a^6 \omega_k^2 
< \hat \phi_k^{i\,2} >_s \right] \nonumber \\
&=& \frac{1}{2a^3} \left[ \frac{\hbar}{2} \left( 
\dot{\rho}_k^2 a^6 + \frac{1}{\rho_k^2} + a^6 \omega^2_k \rho_k^2 \right) + 
<\hat \pip>_s^2 + a^6 \omega_k^2 < \hat \fip >_s^2 \right] \,.
\label{am}
\end{eqnarray}

The adiabatic limit ($\tau \rightarrow -\infty$) for which 
$\rho_k$ is given by $1/ak^{\frac{1}{2}} $ is 
immediately obtained for Eqs.(\ref{inv}), (\ref{indeter}) and 
(\ref{am}) and in particular we note that the uncertainty relation 
(\ref{indeter}) is then minimal and the Fock spaces associated
with {\em a} and {\em b} coincide (see also Eq.(\ref{limad})). 
On the other hand on considering $\tau 
\rightarrow \infty$  one obtains:

\be
<\fip>_s \simeq (2\hbar B)^{\frac{1}{2}} |\alpha| a^{\nu - \frac{3}{2}}
\kap \frac{\g}{\pi} \cos \argom \,
\ee
\be
<\pip>_s \simeq (2\hbar B)^{\frac{1}{2}} |\alpha| H a^{\nu + \frac{3}{2}} 
\left( \nu - \frac{3}{2} \right) \kap
\frac{\g}{\pi} \cos \argom \,
\ee
\be
<\phi_k^{i\,2}>_s \simeq \hbar B a^{2\nu - 3} \quadcap 
\left( \frac{\g}{\pi} \right)^2 \left( \frac{1}{2}
+ 2|\alpha|^2 \cos^2 \argom \right) \,
\ee
\be
<\pi_k^{i\,2}>_s \simeq \hbar B H^2 \left(\nu - \frac{3}{2} \right)^2
a^{2\nu + 3} \quadcap \left( \frac{\g}{\pi} \right)^2 
\left( \frac{1}{2} + 2 |\alpha|^2 \cos^2 \argom \right)
\ee
\be
<(\Delta\pi_k)^2>^{\frac{1}{2}}_s 
<(\Delta\phi_k)^2>^{\frac{1}{2}}_s \simeq \frac{\hbar}{2} B H 
\left(\nu - \frac{3}{2} \right) a^{2\nu} \quadcap 
\left( \frac{\g}{\pi} \right)^2
\ee
\begin{eqnarray}
<H_k>_s &\simeq& \frac{\hbar}{4} B \left( \frac{\g}{\pi} \right)^2 \quadcap 
a^{2\nu} \left[ \tilde{\mu}^2 + H^2 \left( \nu - \frac{3}{2} \right)^2 \right]
\,+ \nonumber \\
&+& \hbar |\alpha|^2 B \left(\frac{\g}{\pi}\right)^2 a^{2\nu}
\quadcap \cos^2 \argom \left[ \left( \nu - \frac{3}{2} \right)^2 H^2 
+ \tilde{\mu}^2 
\right]\,,
\label{asiham}
\end{eqnarray}
with $\tilde{\mu}^2= \mu^2 + 12 \, \xi \,H^2$ 
and in the above the ground state results can be obtained by setting 
$\alpha = 0$. It is immediate to see that $\left( < \phi_k^{i\, 2} >_s \right)
^{\frac{1}{2}}$ with either $\alpha = 0$ or $\alpha \ne 0$, 
that is for the ground state or an arbitrary (classical) state,
has the same behaviour for $\tau \rightarrow \infty$ as $< \phi_k^i >_s$ 
for $\alpha \ne 0$. 
It is for this reason that it is stated that the scalar field 
ground state leads to a classical probability distribution at late times 
\cite{guth}. Further the behaviour of $< \hat H_k >_s$ at late times 
corresponds to particle creation (which can actually oscillates bacause 
of the $\cos^2 \argom$ term \cite{hu}) 
, this is also the reason for the 
increase of the uncertainties in the same limit.  

Let us add a comment on the possible use of a rescaled matter field 
\cite{birrell} $\zeta_k^i = \phi_k^i /a$ in Eq. (\ref{action}). This will lead 
to a modified action:
\be 
S = \sum_{{\kbig},i} S_k^i = \frac{1}{2} \sum_{\kbig} \sum_{i=1,2}
\int a \,d\tau \left[ \dot{\zeta}_k^{i\,2} - \Omega_k^2 \zeta_k^{i\,2} 
- \frac{d}{a\, d\tau} \left(\dot a \zeta_k^{i\,2} \right)\right]
\,,
\label{reaction}
\ee
where $\Omega_k^2=\frac{k^2}{a^2} + \mu^2 + (\xi-\frac{1}{6})\,R$.
The solution to the classical equations of motion in Eq.(\ref{class}) will 
be unchanged however there will be modifications in the subsequent 
Hamiltonian formulation, although the general structure will 
remain unaltered. Thus, for example, the asymptotic behaviour in 
Eq.(\ref{asiham}) will remain the same, but the coefficient multiplying 
it changes. The presence of the {\sl boundary} term in Eq.(\ref{reaction}) 
is important: indeed if it is omitted (which is actually not possible 
with our classical solutions for $\zeta$) the asymptotic behaviour 
in Eq.(\ref{asiham}) is again the same, but coefficient multiplying it 
vanishes for $\mu=0, \xi=\frac{1}{6}$ ({\sl conformal matter}) corresponding 
to no particle creation. 
    
In the above we have illustrated the construction of the vacuum and physical
states for a massive scalar field in a de Sitter space-time through the use
of quantum adiabatic invariants and the associated coherent states.
Such coherent states (in the {\em b} modes) correspond to {\em squeezed}
particle states ({\em a} modes), this is related to particle creation
and is an improvement on the adiabatic approximation in which the classical
limit is obtained for coherent states in the particle modes {\em a}
and there is no particle creation \cite{casadio}. Thus in a
functional Schr\"odinger approach we have obtained
not only the previously suggested
vacuum \cite{bunch} \cite{guth} but since we also have the Fock space
we are able to include particle creation in the classical limit through the
increase in the classical field amplitude.

\end{document}